\documentstyle[a4,amsmath,12pt]{article}

\begin{document}

\titlepage
\begin{flushright}
hep-th/9807055 \\
LPTM 98/66\\
Saclay T98/068
\end{flushright}
\vskip 1cm
\begin{center}
{\bf \Large Superconformal $6D$ (2,0) theories in superspace}
\end{center}
\vskip 1cm
\begin{center}

C. Grojean$^{a}$ {\it and} J. Mourad$^{b,c}$
\end{center} 
\vskip 0.5cm
\begin{center}
$^a$ {\it Service de Physique Th\'eorique, 
CEA-Saclay\\
F-91191 Gif/Yvette Cedex, France}\\
$^b$ {\it Laboratoire de Physique Th\'eorique et Mod\'elisation,\\
Universit\'e de Cergy-Pontoise\\
Site Saint-Martin, F-95032 Cergy-Pontoise, France}\\
$^c$ {\it Laboratoire de Physique Th\'eorique et Hautes Energies
\footnote{Laboratoire associ\'e au CNRS-URA-D0063.},\\
 Universit\'e de Paris-Sud, B\^at. 211, F-91405 Orsay Cedex,
 France}
\end{center}
\vskip 2cm
\begin{center}
{\large Abstract}
\end{center}
\noindent
A geometrical construction of superconformal
transformations in six dimensional (2,0) superspace is proposed.
Superconformal Killing vectors are determined.
It is shown that the transformation of the tensor multiplet
involves a zero curvature non-trivial cochain.
\newpage

\section{Introduction}
\setcounter{equation}{0}
Many reasons motivate the study of
supersymmetric six dimensional chiral theories with sixteen
supercharges \cite{seiberg1}. The more recent one is that the worldvolume
of the five-brane \cite{five,howe,aga,ban,fiveaction} of $M$-theory \cite
{wit}
is described by such a theory.
Not much is known about $M$-theory besides that it contains
membranes and five-branes and that by compactification it reduces
to superstring theory. A recent conjecture by Maldacena
\cite{mald}
states that $M$-theory on $AdS_7\times S^4$ with radii
$R_{sph}=R_{AdS}/2=l_p(\pi N)^{1/3}$ ($l_{p}$ is the eleventh
dimensional Planck length) is dual to the superconformal 
worldvolume
theory describing $N$ coincident five-branes.
Some consequences of this conjecture were examined in \cite{lei}. 
The study of six dimensional (2,0) theories
may thus provide important clues concerning the still mysterious
$M$-theory. Another conjecture on $M$-theory is that of 
Matrix theory \cite{matrix}; here too, (2,0) six dimensional theories
appear from Matrix theory compactified on $T^4$ \cite{roz}.

The aim of this paper is  the geometrical
study of six dimensional (2,0) theories in superspace.
The (2,0) multiplet contains five scalars, one Weyl-symplectic
Majorana spinor and an anti-self dual three form \cite{nahm}.
In section 2, we recall the superfield description
of this multiplet \cite{town1} with one 
superfield in the vector
representation of the $R$-symmetry group $SO(5)$.
This superfield is subject to a constraint which, as shown in
section 2, reproduces the correct multiplet and the equations of
motions. In section 3, we show that there exists an
alternative formulation with the aid of a closed super three-form
which is subject to some constraints. These constraints
are somewhat 
similar to those of the 10D super Yang-Mills constraints
\cite{wit2}
and to the six dimensional (1,0) constraints for the tensorial
multiplet \cite{ber}. For (1,0) six dimensional theories it is
possible to find nontrivial sigma models living
on a quaternionic target space \cite{tow2}. In section 4,
we show that a generalization to sigma models of the
free theory leads only to trivial conformally flat target spaces.
This illustrates the rigidity of (2,0) theories.
In section 5, we define superconformal transformations 
as supercoordinate transformations leaving the super-flat
metric invariant up to a scale. A similar construction for $N$=1
$4D$
theories is considered in \cite{west} and references therein.
We calculate the resulting
super-Killing vectors and show that their Lie algebra
is that of $OSp(6,2|2)$. An algebraic construction
of the superconformal (2,0) algebra was given in 
ref \cite{klaus}, it relies on the triality property
of the six-dimensional conformal group $SO(6,2)$ and results
in the orthosymplectic group $OSp(6,2|2)$. Our geometrical
construction provides a realisation of the generators of 
$OSp(6,2|2)$ in superspace and facilitates the study of
superconformal invariance  in a manifestly supersymmetric
context. In section 6, we determine the transformation
of the scalar superfield under superconformal transformations.
We find that this transformation involves a zero curvature 
1-cochain on
the superconformal Lie algebra. We determine explicitly
this cochain and show that it is non-trivial.
We illustrate the usefulness of the formalism of section 3
by showing that the transformation of the super-three form is
purely geometrical and simpler than that of the scalar superfield.
We collect our conclusions in section 7.
Our notations and some technical results wich 
are used in the text can be
found in the Appendix.

\section{Free supermultiplet}
\setcounter{equation}{0}
The on-shell (2,0) supermultiplet comprises
five scalars $\phi^{i}$ transforming  as a vector of $SO(5)$,
one symplectic Majorana-Weyl fermion and a two-form with self dual
field strength.
The goal of this section is to provide a manifestly supersymmetric
description
of this muliplet \cite{town1}. One may try to consider a scalar superfield 
transforming in the $\bf 5$ of $SO(5)$ whose $\theta=0$ component
is the scalar field. 
Such a superfield would have the general form\footnote{%
In ref \cite{town1} the scalar superfield is in the antisymmetric representation of $Sp(2)$, $\Phi_{[ab]}$. It is related to ours by
$\Phi^i = (\gamma^i)_{ab} \Phi^{[ab]}$.}
\begin{equation}
\Phi^{i}(x,\theta)=\phi^i(x)+\bar\theta\psi^{i}+\dots,
\end{equation}
where $\dots$ stand for terms with two and more $\theta$.
We see that at the one $\theta$ level we have  five independent 
fermions. Since the on-shell supermultiplet has only one fermion
we have to constrain the superfield in such a way that only one
out of the five $\psi^{i}$ be independent. An SO(5) covariant way
of doing this is to impose that
\begin{equation}
\psi^i=\Gamma^i\psi \mbox{ with } \psi={\frac{1}{5}}\Gamma_i\psi^i.
\end{equation}
Note that $\psi^i$ may be written as the $\theta=0$ component of
$D_{\hat \alpha}\Phi^{i}= (\partial_{\hat \alpha}
-(\Gamma^\mu \theta)_{\hat \alpha} \partial_\mu) \Phi^{i}$, so that a manifestly supersymmetric and SO(5)
covariant constraint on the superfield is
\begin{equation}
D\Phi^i={\frac{1}{5}}\Gamma^i\Gamma_j D\Phi^j,
\end{equation}
or equivalently
\begin{equation}
D\Phi^{i}={\frac{1}{4}}\Gamma^{i}_{\ j} D\Phi^{j}.
\end{equation}
When indices are made explicit this constraint reads
\begin{equation}
D_{\alpha a }\Phi^{i}={\frac{1}{4}}(\gamma^{i}_{\ j})
_{ a }^{\  b  }D_{\alpha  b  }\Phi^{j}.\label{cont}
\end{equation}
Our notations can be found in the appendix.
The rest of this section is devoted to the analysis of the
constraint (\ref{cont}). We will show that it reproduces the (2,0)
supermultiplet and the equations of motion.

Let $\Psi_{\alpha  a }=
(\gamma_i)_{ a }^{\  b  }
D_{\alpha  b  }\Phi^{i}/5$ then the supersymmetric
transformation of $\phi^i$ is given by the $\theta=0$ component of
$D\Phi^i=\Gamma^i{\Psi}$.
In order to get the quadratic terms in $\theta$ we
take the $D_{\beta b}$ of the constraint
(\ref{cont}). The following decomposition of the product of two
derivative is useful
\begin{equation}
D_{\alpha  a }D_{\beta  b  }=
-(\gamma^\mu)_{[\alpha \beta]} \Omega_{[ a  b  ]}
\partial_\mu+ \Omega_{[ a   b  ]} D_{(\alpha\beta)}
+(\gamma^i)_{[ a   b  ]} D_{i(\alpha \beta)}
+(\gamma^{ij})_{( a   b  )} D_{[ij][\alpha \beta]},
\label{dec}
\end{equation}
where $(\alpha \beta)$  or $[\alpha \beta]$ means that the quantity
is symmetric or antisymmetric. The quantities appearing in 
(\ref{dec}) are defined , for $N= \emptyset, i$ and $ij$, by
\begin{eqnarray}
D_{N \alpha\beta}=
{\frac{1}{8}} [D_{\alpha a }, D_{\beta  b  }]
(\gamma_N)^{b a}
%
\end{eqnarray}

\noindent
Taking the supersymmetric derivative
of (\ref{cont}) and using the identities
\begin{eqnarray}
&\displaystyle
\gamma^{ij}\gamma^{l}=2\, \eta^{l[j}\gamma^{i]}-
{\frac{1}{2}} \epsilon^{ijl}{}_{mn} \gamma^{mn}
,\\
&\displaystyle
\gamma^{ij}\gamma_{kl} = 
\epsilon^{ij}{}_{klm} \gamma^m
+ 4\, \eta^{[i}{}_{[k} \gamma^{j]}{}_{l]}
+ 2\, \eta^{[i}{}_{[k} \eta^{j]}{}_{l]},
%
\end{eqnarray}
we get the following equations
\begin{eqnarray}
D_{(\alpha\beta)}\Phi^{i}=0,\quad
D_{k(\alpha\beta)}\Phi^{j}={\frac{1}{5}}\eta_{k}{}^{j}
D_{i(\alpha\beta)}\Phi^{i}
\label{DD1},\\
D_{kl[\alpha\beta]}\Phi^{j}=
{\frac{1}{2}}
\left(
\eta_{k}{}^{j} (\gamma^{\mu})_{\alpha\beta} \partial_{\mu}\Phi_{l}-
\eta_{l}{}^{j} (\gamma^{\mu})_{\alpha\beta} \partial_{\mu}\Phi_{k}
\right).
\label{DD2}
\end{eqnarray}
Taking the $\theta=0$ part of these equations shows that
the only new field that appears at this level is the $\theta=0$
component of
\begin{equation}
H_{\alpha\beta}\equiv D_{i(\alpha\beta)}\Phi^{i}.
\end{equation}
Regarded as a matrix, this superfield $H$ can be decomposed in the basis
made by the antisymmetrised products of the Dirac matrices. Taking into account the symmetry and chirality properties ($\alpha$ and $\beta$ are both of the opposed chirality compared to the $\theta$'s), only products of three Dirac matrices appear in this decomposition. So $H$ is equivalent to an anti self-dual three-form given by
\begin{equation}
h_{\mu_1\mu_2\mu_3}=
(\gamma_{\mu_1\mu_2\mu_3})^{\alpha\beta}H_{\alpha\beta}.\label{hh}
\end{equation}
$h_{\mu_1\mu_2\mu_3}$ is anti self-dual because
\begin{equation}
\epsilon_{\mu_1\mu_2\mu_3}{}^{\mu_{4}\mu_5\mu_6}\gamma_{
\mu_{4}\mu_5\mu_6}=-6\, \gamma_{\mu_1\mu_2\mu_3},
\end{equation}
when acting on chiral spinors.
In order to get the transformation of the spinor $\psi$, we
take the supersymmetric derivative of the equation
$ \Psi=\gamma_iD\Phi^i/5$ to get 
\begin{equation}
D_{\beta  b  } \Psi_{\alpha a }=
-(\gamma^{\mu})_{[\beta\alpha]}
(\gamma_{i})_{[ b   a  ]}
\partial_{\mu}\Phi^{i}
+ {\frac{1}{5}}
\Omega_{[ b  a  ]} H_{\beta\alpha}.\label{trf}
\end{equation}
In order to look for possible new fields at the 
level of the product of three $\theta$'s we have to 
calculate the supersymmetric derivative of $H_{\alpha\beta}$.
The consistency of (\ref{trf}) gives
\begin{eqnarray}
&&
2 \, (\Gamma^{\mu})_{\gamma c \beta b}
\partial_\mu \Psi_{\alpha a }
- (\Gamma^{\mu i})_{\beta b \alpha a} (\gamma_i)_{c}{}^{d} 
\partial_\mu \Psi_{\gamma d}
- (\Gamma^{\mu i})_{\gamma c \alpha a} (\gamma_i)_{b}{}^{d} 
\partial_\mu \Psi_{\beta d}
\nonumber\\
&&
-{\frac{1}{5}} \Omega_{ a  b}
D_{\gamma c} H_{\beta\alpha}
-{\frac{1}{5}} \Omega_{ a  c}
D_{\beta b} H_{\gamma\alpha}
=0.
\label{cons}
\end{eqnarray}
Taking the symmetric part in $ a  c$ and
multiplying by $(\Omega^{-1})^{b a }$ we get 
\begin{equation}
D_{\gamma   a }H_{\alpha\beta}=
5 (\gamma^{\mu})_{\gamma\beta} \partial_\mu \Psi_{\alpha a }
+5 (\gamma^{\mu})_{\gamma\alpha} \partial_\mu \Psi_{\beta a }
\label{trh}
\end{equation}
which gives the supersymmetric transformation of the 
three-form and shows that no new degrees of freedom appear at
the three $\theta$'s as well as at higher levels.
Multiplying  (\ref{trh}) by $\cal C^{\alpha\beta}$ gives the
equation of motion of the fermion $\gamma^\mu\partial_\mu\Psi=0$
which when used in (\ref{trf}) gives the bosonic equations
$\partial^\mu\partial_\mu\Phi^i=0$ and 
$\gamma^{\nu}{}_{\gamma}{}^{\alpha}\partial_\nu 
H_{\alpha\beta}=0$.
The simplest way to obtain the latter is to notice that
\begin{equation}
H_{\alpha\beta}=-{\frac{5}{4}}D_{\beta b }\Psi_{\alpha}{}^{b}
\end{equation}
and use the Dirac equation. The equation of motion of
$H_{\alpha\beta}$ reads for the three-form $h$ defined in 
(\ref{hh}) as $dh=d^*h=0$ which gives the Bianchi identity and the 
equations of motion of the three-form.
This equation assures that the three-form $h$ is 
the field strength of a two-form which can be identified with the two-form of the tensoriel supermultiplet.
It remains to prove that equations (\ref{cons}) have no other
content than equations (\ref{trh}). This is easily proved when the
relation 
\begin{equation}
(\gamma^i)_{ a  b  }(\gamma_i)_{c}{}^{d}= 
2\delta_{ a }{}^{d}\Omega_{c  b  }-
2\delta_{ b  }{}^{d}\Omega_{c  a }-\delta_{c}
{}^{d}\Omega_{ a   b  }
\end{equation}
is used. 

\section{Super two-form}
\setcounter{equation}{0}

In this section we formulate the (2,0) free theory 
with a super two-form $B={\frac{1}{2}}B_{MN}E^{M}E^{N}$
 with field strength $H=dB$. 
Here $M=(\mu,\hat\alpha)$,
$E^{\mu}=dx^{\mu}-\bar\theta\Gamma^\mu d\theta$ and
$E^{\hat\alpha}=d\theta^{\hat\alpha}$ are the basis of 1 super-forms
invariant under global supersymmetries.
In a way similar to Yang-Mills theories \cite{wit2}, 
we impose the constraints
\begin{eqnarray}
H_{\hat \alpha \hat \beta \hat \gamma} & = & 0, 
\label{conth}\\
H_{\mu\hat \alpha \hat \beta} & = & (\Gamma_\mu\Gamma_i)_{
\hat \alpha \hat \beta}\tilde\Phi^i,\label{sol1}
\end{eqnarray}
and solve the Bianchi identity $dH=0$. Here
$\tilde \Phi^{i}$ is a scalar superfield which belongs to the
vectorial representation of $SO(5)$. We shall prove that we can
identify $\tilde \Phi^{i}$ with the scalar
superfield $\Phi^{i}$ of the previous section.

\noindent
Terms with four spinorial
indices in the Bianchi identity 
give $H_{\mu(\hat \alpha \hat \beta} (\Gamma^{\mu})_{\hat
\delta \hat \lambda)}=0$ which is satisfied thanks to the
 relation 
\begin{equation}
(\Gamma_{\mu i})_{(\hat\alpha\hat\beta}(\Gamma^\mu)_{\hat
\delta\hat\lambda)}=0,
\end{equation}
which is proved in the Appendix.

\noindent
Terms with three spinorial indices in the Bianchi identity give
\begin{equation}
D_{(\hat \alpha}H_{\hat \beta \hat  \gamma )\mu}
+2H_{\nu\mu(\hat \alpha} (\Gamma^{\nu})_{\hat\beta  \hat \gamma)}
=0.\label{bi}
\end{equation}
Taking into account the Dirac matrices property
\begin{equation}
(\Gamma_{\mu\nu})_{\hat\alpha(\hat\beta}(\Gamma^\mu)_{
\hat\delta\hat\lambda)}+(\Gamma^{i})_{\hat\alpha(\hat\beta}
(\Gamma_{\nu i})_{
\hat\delta\hat\lambda)}=0,
\end{equation}
which is proved in the Appendix,
 the solution to 
(\ref{bi}) can be given with the aid of a  spinorial superfield
$\tilde\Psi_{\hat\alpha}$ as
\begin{eqnarray}
2H_{\nu\mu\hat \beta} & = & 
(\Gamma_{\nu\mu})_{\hat\beta}{}^{\hat\delta}
\tilde\Psi_{\hat\delta},\\
D_{\hat\gamma}H_{\hat\alpha\hat\beta \mu} & = & 
(\Gamma_\mu\Gamma_i)_{\hat \alpha\hat\beta}
(\Gamma^{i})_{\hat\gamma}{}^{\hat\delta}
\tilde\Psi_{\hat \delta}
.\label{sol2}
\end{eqnarray}
Comparing (\ref{sol2}) with (\ref{sol1}) we see that
$\tilde\Phi^i$ and
$\tilde\Psi^{\hat\alpha}$ must be related by
\begin{equation}
(\Gamma^i)_{\hat \alpha}{}^{\hat \beta}
\tilde \Psi_{\hat \beta}=D_{\hat \alpha}\tilde \Phi^i,
\end{equation}
which is equivalent to the constraint (\ref{cont}) obeyed
by the scalar superfield $\Phi^i$ allowing us to identify the two.

\noindent
The terms with two spinorial indices in the Bianchi identity lead
to
\begin{equation}
\partial_{[\mu}
H_{\nu]\hat\alpha\hat\beta}+D_{(\hat\alpha}H_{\hat\beta)\mu\nu}
- H_{\mu\nu\rho} (\Gamma^{\rho})_{\hat\alpha\hat\beta}
=0,
\end{equation}
which is satisfied provided we make the identification
\begin{equation}
H_{\mu\nu\rho}=h_{\mu\nu\rho}\label{idh}
\end{equation}
and use equation (\ref{trf}).

\noindent
The term with one spinorial index is identically zero
du to (\ref{trh}) and  finally the term with no spinorial indices
in the Bianchi identity is zero du to the equation of motion of
$h_{\mu\nu\rho}$ and the identification (\ref{idh}).

\noindent
In brief, the constraints (\ref{conth}) and (\ref{sol1}) 
for the closed super three
form $H$ are equivalent to the constraints (\ref{cont}):
from the superfield $\Phi^i$ we can construct a closed 
super three
form verifying (\ref{conth}) and (\ref{sol1}),
and, conversely, from the constraints
(\ref{conth}) and (\ref{sol1}) on a closed super three-form,
we get a scalar superfield verifying (\ref{cont}).

\section{Sigma model}
\setcounter{equation}{0}

In this section, we search for sigma-model generalizations
of the constraint (\ref{cont}). For (1,0) theories, this
was done in \cite{tow2}.
The five dimensional target
space is assumed to be described by a moving
basis $e^{I}_i(\Phi)$, where $I=1,\dots, 5$ is a flat 
index of $SO(5)$ and $i$ is a curved space index.
We generalise the constraint (\ref{cont}) to have the form
\begin{equation}
e^{I}{}_{i}(\Phi)D_{\alpha  a }\Phi^{i}
=
{\frac{1}{4}}
(\gamma^{I}{}_{J})_{ a }{}^{ b  }
e^{J}{}_{j}(\Phi)D_{\alpha  b  }\Phi^{j}.\label{int}
\end{equation}
{}From (\ref{int}), we can deduce, as in section
2, the transformations of the fields under supersymmetry, their
equations of motions and, in addition, the constraints on the geometry of
the target space.

\noindent
The transformation rule for $\phi^i$ can be deduced from
the $\theta=0$ component of the equation
\begin{equation}
D_{\alpha  a }\Phi^{i}=e_{I}{}^{i}(\gamma^{I})_{ a }{}
^{ b  }
\Psi_{\alpha b  }.
\end{equation}

\noindent
In order to get the constraints on the moving basis 
$e^{I}{}_{i}(\Phi)$,
we
take the spinorial derivative of (\ref{int}) to obtain
\begin{equation}
e^{I}{}_{i}(\Phi)D_{\gamma c}
D_{\alpha  a }\Phi^{i}
-{\frac{1}{4}}
(\gamma^{I}{}_{J})_{ a }{}^{ b  }
e^{J}{}_{j}
(\Phi)D_{\gamma c}D_{\alpha  b  }\Phi^{j}
=f^{I}{}_{\gamma\alpha c  a }
,\label{sif}
\end{equation}
where we used the definition
\begin{equation}
f^{I}{}_{\gamma\alpha c  a }=
\Big(
{\frac{1}{4}}
\gamma^{I}{}_{J}-
\eta^{I}{}_{J}
\Big)_{ a }{}^{ b  }
\ e^{J}{}_{j,k}D_{\gamma c}\Phi^{k}
D_{\alpha  b  }\Phi^{j}.\label{ff}
\end{equation}
It is convenient, for the analysis of (\ref{sif}), 
to use the following decomposition 
on the $SO(5)$ gamma matrices
\begin{equation}
f^{I}{}_{\gamma\alpha c  a }=
f^{I}{}_{\gamma\alpha} \Omega_{c  a }
+f^{I}{}_{J\gamma\alpha} (\gamma^J){}_{c  a }
+f^{I}{}_{JK\gamma\alpha} (\gamma^{JK}){}_{c  a }.
\end{equation}
Then equation (\ref{sif}), using the decomposition (\ref{dec}),
gives the following relations which replace eqs (\ref{DD1}) and 
(\ref{DD2})
\begin{eqnarray}
e^{I}{}_{j} D_{(\gamma\alpha)} \Phi^{j}
& = & f^{I}{}_{(\gamma\alpha)}
,\\
e_{Ij} D_{K(\gamma\alpha)}\Phi^{j} & = &
{\frac{1}{5}}
\eta_{IK} e^{J}{}_{j} D_{J(\gamma\alpha)} \Phi^j
+{\frac{16}{15}}
f_{IK(\gamma\alpha)}
- {\frac{4}{15}} 
f_{KI(\gamma\alpha)}
,\\
e_{Ij} D_{MN[\gamma\alpha]} \Phi^{j} & = &
{\frac{1}{2}}
\left( 
\eta_{MI} e_{Nj} (\gamma^\mu)_{\gamma\alpha}
\partial_{\mu} \Phi^{j}
-\eta_{NI} e_{Mj} (\gamma^\mu)_{\gamma\alpha}
\partial_{\mu} \Phi^{j} \right)
\\
&&
-{\frac{1}{5}}
\epsilon^{JK}{}_{IMN} f_{JK[\gamma\alpha]}
 + {\frac{2}{5}}(\eta_{MI} f_{N[\gamma\alpha]}-
\eta_{NI} f_{M[\gamma\alpha]})
+{\frac{4}{5}}
f_{IMN[\gamma\alpha]},
\nonumber
\end{eqnarray}
as well as the following constraints on 
$f^{I}{}_{\gamma\alpha  c  a }$:
\begin{eqnarray}
&&
f^{I}{}_{I\gamma\alpha}=0,\quad
f_{(IK)[\gamma\alpha]}=0
,\label{cont1}\\
&&
f_{I[\gamma\alpha]}=2f^{J}{}_{JI[\gamma\alpha]}
,\quad
\epsilon^{IJKMN}f_{KMN[\gamma\alpha]}=-2f^{IJ}{}_{[\gamma\alpha]}
,\\
&&
f^{I}{}_{MN(\gamma\alpha)}=
{\frac{1}{8}}
\left(
-\eta^{I}{}_{N} f_{M(\gamma\alpha)}+
\eta^{I}{}_{M}f_{N(\gamma\alpha)}
\right)
-{\frac{1}{6}}
\epsilon^{IJK}{}_{MN} f_{[JK](\gamma\alpha)}
.\label{cont3}
\end{eqnarray}
The structure of 
$f^{I}{}_{\gamma\alpha c  a }$ given in (\ref{ff}) 
implies a number of
constraints which are identically satisfied. These are
given by
\begin{equation}
f^{I}{}_{\gamma\alpha}
=2 f^{J}{}_{J}{}^{I}{}_{\gamma\alpha}
,\quad
f^{J}{}_{J\gamma\alpha}=0
,\quad 
6 f_{[IMN]\gamma\alpha}=
-\epsilon_{IMN}{}^{JK} f_{JK\gamma\alpha}.
\end{equation}
So only the second constraint in (\ref{cont1})
and (\ref{cont3}) are not identically
satisfied. It turns out that these two imply that the 
moving basis must be such
\begin{equation}
de^I \wedge e^J
+ de^J \wedge e^I
-{\frac{2}{5}} \eta^{IJ} \eta_{KL} de^K \wedge e^L =0
.\label{cond}
\end{equation}
It is possible to solve equations (\ref{cond}) to
get $e^I=\Sigma d \Upsilon^{I}$ for some functions 
$\Sigma$ and $\Upsilon^I$ which means that the target space
is conformally flat. By a change of coordinates in the target
space $\Phi^i\rightarrow \Upsilon^I$ the theory is transformed to the
free theory of section 2. So no non-trivial sigma-models are
allowed by (2,0) supersymmetry.

\section{Superconformal transformations}
\setcounter{equation}{0}
In this section we give a geometrical construction of the
superconformal transformations and determine explicitly the
realisation of the generators in the (2,0) superspace.

The flat supersymmetric metric in superspace is given by
\begin{equation}
g=\eta_{\mu\nu}E^{\mu}\otimes E^\nu.
\end{equation}
Notice that the other term appearing {\it a priori} in the
$(2,2)$ superspace, 
$C_{\hat \alpha \hat \beta} E^{\hat \alpha} \otimes E^{\hat \beta}$, is forbidden by chirality in $(2,0)$ theories. 
A supercoordinate transformation is generated by an even  vector field
$\xi=\xi^{\mu} E_{\mu}+\xi^{\hat\beta}E_{\hat\beta}$,
where $\xi^\mu$ is even and $\xi^{\hat\beta}$ is odd.
Under this transformation, the supercoordinates $Z^M$ 
transform as
\begin{eqnarray}
\delta x^{\mu} & = & \xi^{\mu}+\theta^{\hat\alpha}
(\Gamma^\mu)_{\hat\alpha\hat\beta}\xi^{\hat\beta}
,\nonumber\\
\delta \theta^{\hat\alpha} & = & \xi^{\hat\alpha},
\end{eqnarray}
and the metric varies as
\begin{equation}
\delta g=L_{\xi}g,
\end{equation}
where $L_{\xi}$ is the Lie derivative with respect to the vector
$\xi$. A superconformal transformation is defined by the
requirement that the transformed metric be proportional
to the initial one, that is
\begin{equation}
\delta g 
= \alpha(Z) g
,\label{su}
\end{equation}
where $\alpha$ is {\it a priori} an arbitrary superfield. 
The vector field
$\xi$ is said to be a superconformal Killing vector.
The use of the relation $[L_{\xi},L_{\xi'}]
=L_{[\xi,\xi']}$,
where $[,]$ is the Lie bracket, shows that if $\xi$ and
$\xi'$ are two superconformal Killing vector fields with
scales $\alpha$ and $\alpha'$ then $[\xi,\xi']$ is also a 
superconformal Killing vector with scale
$\xi(\alpha')-\xi'(\alpha)$, so that the set of all
$\xi'$s forms a Lie algebra.
 
\noindent
In order to determine explicitly the superconformal Killing
vectors we first calculate the 
 variation of the basis super one-forms:
\begin{eqnarray}
\delta
E^{\mu}=(d\iota_{\xi}+\iota_{\xi}d)E^{\mu}
= (\partial_{\nu} \xi^{\mu}) E^{\nu}
+(D_{\hat\alpha}\xi^{\mu} + 
2 (\Gamma^{\mu}){}_{\hat\alpha\hat\beta}\xi^{\hat \beta})
E^{\hat\alpha},
\end{eqnarray}
so the condition (\ref{su}) is verified provided
\begin{eqnarray}
D_{\hat\alpha}\xi^{\mu}
+ 2 (\Gamma^{\mu}){}_{\hat\alpha\hat\beta}
\xi^{\hat\beta}=0
,\label{su1}\\
\partial_{\mu}\xi_{\nu}+\partial_{\nu}\xi_{\mu}=
\alpha\eta_{\mu\nu}\label{su2}.
\end{eqnarray}
We shall show that the equation (\ref{su2}) is
a consequence of (\ref{su1}), so that the solutions of the
latter 
determine all superconformal Killing vectors.
Note that, by equation (\ref{su1}), $\xi^{\hat \beta}$ is 
determined in
terms of $\xi^{\mu}$ by
\begin{equation}
\xi^{\hat\beta}=
-{\frac{1}{12}} (\Gamma^{\mu}){}^{\hat \beta\hat\alpha}
D_{\hat\alpha}\xi_{\mu}\label{xx}.
\end{equation}
This allows to write equation (\ref{su1}) in terms of $\xi_{\mu}$
as
\begin{equation}
D_{\hat\alpha}\xi^{\mu}=
{\frac{1}{6}}
(\Gamma^{\mu\nu}){}_{\hat\alpha}{}^{\hat\beta}
D_{\hat\beta}\xi_{\nu} 
\label{su3},
\end{equation}
which is very similar to the constraint of the scalar superfield (\ref{cont}),
the vectorial structure in $SO(5)$ being replaced by the same one in 
$SO(1,5)$.
In order to analyse equation (\ref{su3}) it is convenient
to decompose the product of two spinorial derivatives as
\begin{eqnarray}
D_{\hat\alpha}D_{\hat\beta} & = & 
-(\Gamma^\nu)_{(\hat\alpha\hat\beta)}
\partial_{\nu}+(\Gamma^{\nu ij})_{[\hat\alpha\hat\beta]}
D_{\nu[ij]}
\nonumber\\
&&+(\Gamma^{\mu_1\mu_2\mu_3})_{[\hat\alpha\hat\beta]}
D_{\mu_1\mu_2\mu_3}+
(\Gamma^{\mu_1\mu_2\mu_3 i})_{[\hat\alpha\hat\beta]}
D_{\mu_1\mu_2\mu_3 i},
\end{eqnarray}
then taking the spinorial derivative of (\ref{su3})
yields 
\begin{eqnarray}
&&
\partial_{\mu}\xi_{\nu}+\partial_{\nu}\xi_{\mu}=
{\frac{1}{3}}\partial_{\rho}\xi^{\rho}\eta_{\mu\nu}
,\label{ss1}\\
&&
D_{\nu[ij]}\xi_{\mu}={\frac{1}{6}}D_{\rho[ij]}\xi^{\rho}
\eta_{\nu\mu}
,\label{ss2}\\
&&
D_{\mu_1\mu_2\mu_3}\xi_{\mu_4}=
-{\frac{1}{4}}
\left(
\eta_{\mu_4[\mu_1}\partial_{\mu_2}\xi_{\mu_3]}
-{\frac{1}{6}}\epsilon_{\mu_1\mu_2\mu_3}{}^{\nu_1\nu_2\nu_3}
\eta_{\mu_4[\nu_1}\partial_{\nu_2}\xi_{\nu_3]}
\right)
\label{ss3},\\
&&
D_{\mu_1\mu_2\mu_3 i}\xi_{\mu_4}=0\label{ss4}.
\end{eqnarray}
The first equation is equivalent to (\ref{su2}) which
shows that (\ref{su2}) is contained in (\ref{su1}).
Let $\zeta_{\mu}(x)$, $\zeta^{\hat\beta}(x)$ and
$\zeta^{ij}(x)$ be the $\theta=0$ components of
$\xi_{\mu}$, $\xi^{\hat\beta}$ and $D_{\rho[ij]}\xi^{\rho}$,
then equations (\ref{ss1}, \ref{ss2}, \ref{ss3}) and (\ref{ss4})
 show that the superfield $\xi_{\mu}$ is
determined in terms of the $\zeta$'s. These are solutions to
the following decoupled equations which are consequences of
(\ref{ss1})
\begin{eqnarray}
&&
\partial_{\mu}\zeta_{\nu}+\partial_{\nu}\zeta_{\mu}=
{\frac{1}{3}}
\partial_{\rho}\zeta^{\rho}\eta_{\mu\nu}
,\label{rr1}\\
&&
(\Gamma_{\mu}){}_{\hat\alpha\hat\beta} \partial_{\nu} \zeta^{\hat\beta}
+
(\Gamma_{\nu}){}_{\hat\alpha\hat\beta} \partial_{\mu} \zeta^{\hat\beta}=
{\frac{1}{3}}
\eta_{\mu\nu}
(\Gamma_{\rho}){}_{\hat\alpha\hat\beta} 
\partial^{\rho} \zeta^{\hat\beta}
,\label{rr2}\\
&&
\partial_{\mu} \zeta^{ij}=0
\label{rr3}.
\end{eqnarray}
The solutions to (\ref{rr1}) are the well-known conformal Killing
vectors
\begin{equation}
\zeta_\mu=a_\mu\ +\ a_{[\mu\nu]}x^\nu\ +\ \lambda x_\mu\ +\ 
(x^2\eta_{\mu\nu}-2x_\mu
x_\nu)k^{\nu}
,\label{sro1}
\end{equation}
where  $a_\mu,\ a_{\mu\nu},\ \lambda$ and $k_\mu$ are
parameters of infinitesimal translations, Lorentz transformations,
dilatations and special conformal transformations.
Similarly, 
the solutions to (\ref{rr2}) are determined in terms of two
constant spinors $\epsilon$ and $\eta$ 
(respectively simplectic-Majorana-Weyl and anti-simplectic-Majorana-Weyl spinors)
as
\begin{equation}
\zeta^{\hat\beta}=\epsilon^{\hat\beta}
+ x_\mu
(\Gamma^{\mu}){}^{\hat\beta}{}_{\hat\alpha}\eta^{\hat\alpha},
\label{sro2}
\end{equation}
$\epsilon$ is the parameter of a supersymmetry transformation 
and $\eta$ that of a special supersymmetry transformation.
Finally the solution of (\ref{rr3}) is given by
\begin{equation}
\zeta^{ij}={\frac{1}{4}}\epsilon^{[ij]},\label{sro3}
\end{equation}
where $\epsilon^{[ij]}$ are constants which represent
infinitesimal $SO(5)$ rotations.
The complete $\theta$ expansion of the superfield $\xi_{\mu}$
follows from the  solutions
(\ref{sro1}, \ref{sro2}, \ref{sro3}) and from the equations 
(\ref{ss1}, \ref{ss2}, \ref{ss3})
after some tedious algebra as
\begin{eqnarray}
\xi^{\mu} & = & 
\zeta^\mu -
2\bar\theta\Gamma^{\mu}
\zeta+\bar\theta
\left(
\Gamma^{\mu i j}\zeta_{ij} 
+{\frac{1}{4}}\,
\Gamma^{\mu\mu_1\mu_2}\partial_{\mu_1}
\zeta_{\mu_2}
\right) \theta
\nonumber\\
&&+{\frac{1}{2}}
\, \bar\theta\Gamma^{\mu\mu_1\mu_2}\theta\
\bar\theta\Gamma_{\mu_1}\partial_{\mu_2}\zeta
-{\frac{1}{64}}\,
\bar\theta\Gamma^{\mu\mu_1\mu_2}\theta\
\bar\theta \Gamma_{\mu_1\mu_2\mu_3}\theta\
\partial_{\rho}\partial^{\rho}\zeta^{\mu_3}.\label{xim}
\end{eqnarray}
In order to have the complete expression of the 
superconformal Killing vector field $\xi$ we have to calculate
$\xi^{\hat\alpha}$ from equation (\ref{xx}), the result, after
some arrangements is 
\begin{eqnarray}
\xi^{\hat\alpha}
& = & 
\left[
\zeta-\Gamma_{ij}\theta\ \zeta^{ij} 
+{\frac{1}{12}}
\theta\ \partial_\mu\zeta^\mu
-{\frac{1}{4}}
\Gamma^{\mu\nu}\theta\ \partial_{\mu}\zeta_{\nu}
\right.\nonumber\\
&&
-{\frac{1}{6}}
\theta\ \bar\theta\Gamma^\mu\partial_\mu\zeta
-{\frac{1}{2}} \Gamma^{\mu\nu}\theta\ 
\bar\theta\Gamma_\mu \partial_\nu\zeta
-{\frac{1}{24}}
\Gamma_{\mu_1\mu_2}\partial_{\mu_3}\zeta\
\bar\theta\Gamma^{\mu_1\mu_2\mu_3}
\theta\nonumber\\
&&\left.
+{\frac{1}{32}}
\Gamma^{\mu_1\mu_2}\theta\
\bar\theta \Gamma_{\mu_1\mu_2\mu_3}\theta\
\partial_{\sigma}\partial^{\sigma}\zeta^{\mu_3}
\right]^{\hat\alpha}
.\label{xia}
\end{eqnarray}
It is also possible to determine the scale $\alpha$,
\begin{eqnarray}
\alpha={\frac{1}{3}}
\left[\partial_\mu\zeta^\mu - 2\bar \theta\Gamma^{\mu}
\partial_\mu\zeta\right].
\end{eqnarray}
The Lie algebra of superconformal transformations
can be deduced from the calculation of the Lie bracket of
two super-Killing vectors $\xi_1$ and $\xi_2$ which we denote by
$\xi_3$.
Let the parameters determining the $\xi_a$ ($a=1,2,3$) be given by
$a^{\mu}_a,\ a_{a}^{\mu}{}_{\nu},\ \lambda_a,\ k^{\mu}_{a},\  
\epsilon_a,\  \eta_a$ and $\epsilon^{ij}_{a}$ then the parameters of
$\xi_3$ are given by
\begin{eqnarray}
a_3^\mu & = &  
a_1^\nu 
a_2^{\mu}{}_{\nu}-a_2^{\nu}a_1^{\mu}{}_{\nu}+\lambda_2a_1^\mu
-\lambda_1a_2^\mu
- 2 {\bar \epsilon}_1\Gamma^\mu\epsilon_2,
\nonumber\\
a_3^{\mu\nu} & = & 
2\ (a_1^\nu k_2^\mu-a_1^\mu k_2^\nu-a_2^\nu
k_1^\mu+a_2^\mu k_1^\nu+\bar\epsilon_2\Gamma^{\mu\nu}\eta_1
-\bar\epsilon_1\Gamma^{\mu\nu}\eta_2)
\nonumber\\
&&+a_1^{\rho\nu}a_2^{\mu}{}_{\rho}
-a_2^{\rho\nu}a_1^{\mu}{}_{\rho}
,\nonumber\\
\lambda_3 & = & 
-2\ (a_1^\nu k_{2\nu}-a_2^\nu k_{1\nu}
+\bar\epsilon_1\eta_2- {\bar\epsilon}_2 \eta_1)
,\nonumber\\
k_3^\mu & = & 
a_2^{\mu\nu}k_{1\nu}-a_1^{\mu\nu}k_{2\nu}+
\lambda_1 k_2^\mu -\lambda_2 k_1^\mu 
-2\bar\eta_1\Gamma^\mu\eta_2
,\nonumber\\
\epsilon_3 & = & 
a_{1\mu} \Gamma^\mu\eta_2-a_{2\mu} \Gamma^\mu\eta_1
+{\frac{1}{4}}(\epsilon^{ij}_1\Gamma_{ij}\epsilon_2-
\epsilon^{ij}_2\Gamma_{ij}\epsilon_1)-{\frac{1}{2}}(
\lambda_1\epsilon_2-\lambda_2\epsilon_1)
\nonumber\\
&&
-{\frac{1}{4}}(a_{1\mu\nu}\Gamma^{\mu\nu}\epsilon_2-
a_{2\mu\nu}\Gamma^{\mu\nu}\epsilon_1)
,\nonumber\\
\eta_3 & =& 
k_{1\mu}\Gamma^\mu\epsilon_2-k_{2\mu}\Gamma^\mu\epsilon_1
+{\frac{1}{4}}(\epsilon_1^{ij}\Gamma_{ij}\eta_2-
\epsilon_2^{ij}\Gamma_{ij}\eta_1)+
{\frac{1}{2}}(\lambda_1\eta_2-\lambda_2\eta_1)
\nonumber\\
&&
-{\frac{1}{4}}(a_{1\mu\nu}\Gamma^{\mu\nu}\eta_2-
a_{2\mu\nu}\Gamma^{\mu\nu}\eta_1)
,\nonumber\\
\epsilon^{ij}_3 & = & 
\epsilon_2^{ik}\epsilon_1^{j}{}_{k}
-\epsilon_1^{ik}\epsilon_2^{j}{}_{k}-4(\bar\epsilon_1
\Gamma^{ij}\eta_2-\bar\epsilon_2
\Gamma^{ij}\eta_1)
.\label{li}
\end{eqnarray}
These relations encode the Lie algebra of superconformal
transformations. It can be readily
verified that this Lie algebra is that
of $OSp(6,2|2)$ given in reference \cite{klaus}.

\section{Superconformal invariance}

\setcounter{equation}{0}
The vector fields $\xi$, in general, do not
commute with $D_{\hat\alpha}$; using the relation (\ref{su1}) on the Killing vector, the precise commutation
relations is given by
\begin{equation}
[D_{\hat\alpha},\xi]=(D_{\hat\alpha}\xi^{\hat\beta})D_{\hat\beta}.
\label{commu}
\end{equation}
This equation shows that under a superconformal transformation
the vector fields $D_{\hat\alpha}$ are not invariant. However they
transform in such a way as not to mix with the 
$\partial_\mu$. This could have been also a starting point for
the definition of superconformal transformations. For future use
it is important to explicit the right hand side of (\ref{commu})
as
\begin{equation}
D_{\hat\alpha} \xi_{\hat\beta}=
-{\frac{1}{4}}
(\Gamma^{\mu\nu})_{\hat\beta\hat\alpha} \partial_\mu\xi_\nu
+{\frac{1}{12}}
C_{\hat\beta\hat\alpha} \partial_\sigma\xi^\sigma
+{\frac{1}{12}}
(\Gamma^{ij})_{\hat\beta\hat\alpha}
D_{\sigma[ij]} \xi^{\sigma}
,\label{dxi}
\end{equation}
where $D_{\sigma[ij]}
\xi^{\sigma}$ is given, from (\ref{xim}), by
\begin{equation}
D_{\sigma[ij]}\xi^{\sigma}=
-12\zeta_{ij}+2\bar\theta\Gamma_{ij}\Gamma^{\sigma}
\partial_\sigma\zeta-
{\frac{3}{4}}
\bar\theta\Gamma_\nu\Gamma_{ij}\theta
\ \partial_\sigma\partial^\sigma\zeta^\nu
.\label{dsi}
\end{equation}
{}From equation (\ref{commu}) we can easily deduce that the
constraint (\ref{cont}) is not invariant under the
transformations $\delta\Phi^i=\xi(\Phi^i)$. 
This motivates the introduction of
a connection $\Lambda^{i}{}_{j}(\xi)$ so that the transformation
of $\Phi^i$ becomes
\begin{equation}
\delta_\xi\Phi^i=\xi(\Phi^i)+\Lambda^{i}{}_{j}(\xi)\Phi^j.
\end{equation}
We shall explicitly construct $\Lambda(\xi)$ later.
Here, we examine some of its mathematical properties.
In fact $\Lambda$ is not strictly a connection because in general
we do not have $\Lambda(f\xi)=f\Lambda(\xi)$, it is only
a cochain on the Lie algebra of superconformal transformations
realised with the superconformal Killing vectors. It is valued in 
the tensor product of the algebra of superfields  and the algebra
of $5\times 5$ matrices.
This cochain is
 however not arbitrary.
The requirement $[\delta_{\xi},\delta_{\xi'}]=\delta_{[\xi,\xi']}$
gives the following consistency condition on $\Lambda(\xi)$
\begin{equation}
\xi(\Lambda(\xi'))-\xi'(\Lambda(\xi))-
\Lambda([\xi,\xi'])+[\Lambda(\xi),\Lambda(\xi')]=0,\label{che}
\end{equation}
where we have used a matrix notation for $\Lambda$.
In order to make the algebraic meaning of (\ref{che})
clearer, recall that given a n-cochain on a Lie algebra, 
${\it
g}$, $\alpha(g_1,\dots,g_n)$ which is valued in a 
${\it g}$-module ${\cal M}$ then the Chevalley exterior 
derivative ${\bf s}\alpha$ is a n+1-cochain given by
\begin{eqnarray}
{\bf s}\alpha(g_1,\dots,g_{n+1})
=
\sum_{1 \leq i \leq n+1} (-1)^{i-1} 
g_i \alpha (g_1 \dots {\hat g}_{i} \ldots g_{n+1})
\nonumber\\
+ \sum_{1\leq i < j \leq n+1 } {(-1)^{i+j}}
\alpha([g_i,g_j] g_1 \dots {\hat g}_{i} \dots {\hat g}_{j} \dots g_{n+1} ).
\end{eqnarray}
where the notation ${\hat g}_{i}$ means that the element $g_i$ 
has been omited.  
The Chevalley exterior derivative verifies ${\bf s}^2=0$
and the graded Leibniz rule. This allows to write
condition (\ref{che}) as
\begin{equation}
{\bf s}\Lambda+\Lambda^2=0,\label{chev}
\end{equation}
where the exterior product of two cochains is defined in a way
similar to the exterior product of two forms.
The algebraic interpretation of condition (\ref{che})
is thus that  $\Lambda$ has vanishing curvature.
A particular solution to (\ref{chev}) is given by a ``pure gauge"
\begin{equation}
\Lambda=\beta^{-1}{\bf s}\beta,\label{cheva}
\end{equation}
where $\beta$ is a 0-cochain.
Equation (\ref{cheva})
reads explicitly
\begin{equation}
\Lambda(\xi)=\beta^{-1}\xi(\beta).\label{pg}
\end{equation}   
Note that if $\Lambda$ is a pure gauge then
if we define $\Phi'$ by $\beta\Phi$
we get $\delta\Phi'=\xi(\Phi')$, so that a rescaling   
of the superfield allows the elimination of $\Lambda$.
We shall show below that the correct 
$\Lambda$ is not a pure gauge but has vanishing curvature.

In order to explicitly determine $\Lambda(\xi)$,
we demand that if $\Phi$ verifies the constraint (\ref{cont})
then so does $\delta\Phi=\xi(\Phi)+\Lambda(\xi)\Phi$.
This is true provided
\begin{eqnarray}
& \displaystyle
D_{\hat\alpha} \Lambda^{i}{}_{j}
-{\frac{1}{4}}
(\Gamma^{i}{}_{k})_{\hat\alpha}{}^{\hat\beta}
D_{\hat\beta}\Lambda^{k}{}_{j}=0
,\label{un}\\
& \displaystyle
(\gamma^{i}{}_{j})_{ a }{}^{ b  }
D_{\alpha b  }\xi^{\beta c}-
D_{\alpha a }\xi^{\beta b  }
(\gamma^{i}{}_{j})_{ b  }{}^{c}
-
\left(
(\gamma^{i}{}_{k})_{ a }{}^{c}
\Lambda^{k}{}_{j}-\Lambda^{i}{}_{k}
(\gamma^{k}{}_{j})_{ a }{}^{c}
\right)
{\delta}_{\alpha}{}^{\beta}=0
.\label{deu}
\end{eqnarray}
By replacing $D_{\hat\alpha}\xi_{\hat\beta}$
in   equation (\ref{deu}) by its expression (\ref{dxi})
 we determine
$\Lambda_{ij}$ up to an arbitrary function $\chi$ as
\begin{equation}
\Lambda_{ij}=-{\frac{1}{3}}
D_{\nu[ij]}\xi^{\nu}+\eta_{ij} \chi.
\end{equation}
This expression shows that $\Lambda$ is actually valued in the
Lie algebra of $u(1)\oplus so(5)$ rather than $gl_5$.
The function $\chi$ is calculated with the aid of equation 
(\ref{un}) after using the explicit expression of
$D_{\nu[ij]}\xi^{\nu}$ given in (\ref{dsi}) as well as its spinorial
derivative which is given by
\begin{equation}
D_{\hat\alpha}D_{\sigma[ij]}\xi^{\sigma} = 2
(\Gamma^\nu\Gamma_{ij})_{\hat\alpha\hat\delta}
\partial_\nu\xi^{\hat\delta}.
\end{equation}
The resulting expression for $\chi$ turns out to be very simple 
\begin{equation}
\chi={\frac{1}{3}}\partial_\mu\xi^\mu+\lambda'= \alpha+\lambda',
\label{f}
\end{equation}
where $\lambda'$ is an arbitrary constant which has to
be set to zero in order to have $\Lambda(0)=0$.
Finally, we are in position to deduce the complete expression
for the 1-cochain $\Lambda_{ij}$ as
\begin{eqnarray}
\Lambda_{ij}(\xi) = 
4\zeta_{ij}-{\frac{2}{3}}
\bar\theta\Gamma_i\Gamma_j\Gamma^\sigma \partial_\sigma\zeta
+{\frac{1}{4}}
\bar\theta\Gamma_\nu\Gamma_{ij}\theta
\partial_\sigma\partial^\sigma\zeta^\nu
+{\frac{1}{3}}\eta_{ij}
\partial_\mu\zeta^\mu.
\end{eqnarray}
The above expression for $\Lambda$ shows clearly that it
is a non-trivial zero-curvature 1-cochain because it cannot be
written as 
 a ``pure gauge"-cochain of the form (\ref{pg}).
It remains to check that $\Lambda$ indeed verifies the 
consistency condition (\ref{chev}). Note first that the $u(1)$
part of (\ref{chev}) reads 
\begin{equation}
{\bf s}\chi=0,
\end{equation}
which is true because $\chi$ is given by equation (\ref{f})
and $\alpha$ is closed : 
$\alpha([\xi,\xi'])= \xi(\alpha(\xi'))-\xi'(\alpha(\xi))$.
The antisymmetric part of $\Lambda$ has also a 
vanishing curvature. This can be verified by a direct calculation
using (\ref{li}) but a more simple way is to examine
the transformation property of the super three-form.

\noindent
We shall show that the simple geometrical transformation
\begin{equation}
\delta H=L_{\xi} H\label{ttt}
\end{equation} 
leaves the constraints (\ref{conth}) and (\ref{sol1}) invariant.
Note first that the transformation of the moving basis
$E^\mu$ does not involve the spinorial basis $E^{\hat\alpha}$,
so the constraint (\ref{conth}) is left invariant by (\ref{ttt}).
The transformation of the $\mu\hat\alpha\hat\beta$
components of $H$ under (\ref{ttt})
reads
\begin{eqnarray}
\delta H_{\mu\hat\alpha\hat\beta}
=
\xi(H_{\mu\hat\alpha\hat\beta})
+H_{\nu\hat\alpha\hat\beta}\partial_\mu\xi^\nu
+H_{\mu\hat\alpha\hat\delta}D_{\hat\beta}\xi^{\hat\delta}
+H_{\mu\hat\beta\hat\delta}D_{\hat\alpha}\xi^{\hat\delta}.
\end{eqnarray}
The use of relation (\ref{dxi}) allows to write
\begin{equation}
\delta H_{\mu\hat\alpha\hat\beta}=
(\Gamma_\mu\Gamma_i)_{\hat\alpha\hat\beta}\delta\Phi^{i},
\label{tth}
\end{equation}
with $\delta\Phi^i$ given by
\begin{equation}
\delta\Phi^i=\xi(\Phi^i)
+{\frac{1}{3}}
\left(
\partial_\mu\xi^\mu\Phi^i
- D_{\sigma}{}^{i}{}_{j} \xi^\sigma \Phi^{j}
\right).
\end{equation}
Relation (\ref{tth}) shows that 
the constraint (\ref{sol1}) is left invariant by the
transformation (\ref{ttt}). In addition we can identify the
transformation of $\Phi^i$ which agrees, as it should,
 with the previously
calculated one.
This geometrical construction of $\Lambda$ assures the vanishing curvature property du to the Lie structure of the transformations of the three-form $H$.

\section{Conclusion}

We have given two equivalent
formulations of the tensorial multiplet of $6D$ (2,0) theories.
We have illustrated the rigidity of (2,0) supersymmetries
by showing the triviality of the sigma-model.
We note here that this triviality can be understood from the super
three-form point of view; indeed, the generalization of the 
constraints (\ref{sol1}) to a curved target space manifold,
$H_{\mu{\hat \alpha}{\hat \beta}}=
(\Gamma_{\mu I})_{\hat\alpha\hat\beta}\, {e^{I}}_{i} {e^{i}}_{J}
\Phi^J$  does not lead to an interacting sigma-model.  

\noindent
Note that the five scalars $\phi^i$ describe the transverse 
fluctuations of the five-brane. The constraint (\ref{cont})
coincides with the equation obtained in \cite{howe}  \
using the superembedding formalism applied for a five-brane
in a super-flat background, in the physical gauge and
in  the linearised approximation. This suggests the existence 
of a formulation of the full non-linear equations of motion of the
five-brane which is analogous to the one presented in section 3
and which is based on constraints on a three-form in a non
super-flat background. We hope to come back to this issue 
in more details elsewhere.

\noindent
We have also realised the superconformal transformations as 
derivations
in superspace. This gives a geometric construction 
of the superconformal Lie algebra compared to the algebraic one
presented in \cite{klaus}.
Moreover, this allows to realise, in a
manifestly supersymmetric way, the transformation
of the supermultiplet. This gives an alternative proof to
the one relying on the component formalism 
presented in \cite{klaus}
that the linearised equations of motion of the
five-brane are superconformally invariant.

We have found that the transformation
of the three-form involves only coordinate reparametrisations
whereas, for the scalar superfield, the introduction
of a cochain is in addition needed. 
This suggests that the formulation in terms of super three-forms
may be helpful in the construction of superconformal interacting
theories and shed some light on the conjecture 
\cite{mald} relating them to
$M$-theory.
In this respect, recently it has been shown \cite{kallosh}
that the bosonic action for
$M$-theory five-branes in their near horizon background
have a non-linearly realised conformal invariance.
Our results may be useful in extending this invariance to a 
superconformal one.

\appendix
\section{Conventions}

\setcounter{equation}{0}

In this Appendix we collect our conventions and notations.
The 6D (2,0) supersymmetry algebra is conveniently described
as a chiral truncation of the reduction of the 11D algebra.
The 11D superalgebra reads
\begin{equation}
\{Q_{\hat \alpha},Q_{\hat \beta}\}= 2(\Gamma^{\hat \mu}C)_{\hat
\alpha \hat\beta}P_{\hat\mu},\label{onze}
\end{equation}
where $\hat\alpha=1,\dots, 32,\ \hat\mu=0,\dots 10$ and
$C_{\hat\alpha\hat\beta}$ is an 
antisymmetric matrix verifying
\begin{equation}
C^{-1}\Gamma^{\hat\mu}C=-\Gamma^{\hat\mu T}.
\end{equation}
The reality condition on 11D fermions reads
\begin{equation}
\Psi=C\bar\Psi^{T},
\end{equation}
or equivalently
\begin{equation}
\bar\Psi^{\hat\alpha}=C^{\hat\alpha\hat\beta}\Psi_{\hat\beta}
\equiv \Psi^{\hat\alpha},
\end{equation}
where $C^{\hat\alpha\hat\beta}$ is the inverse of 
$C_{\hat\alpha\hat\beta}$.  We shall use $C$ to raise
and lower indices and the notation
$(\Gamma^\mu)_{\hat\alpha\hat\beta}$ for  
$(\Gamma^\mu C)_{\hat\alpha\hat\beta}$. 
Under reduction to six dimensions
the spinorial index $\hat\alpha$ decomposes as
$\alpha  a $ where $\alpha$ is an $SO(5,1)$ spinorial
index and $ a $ an $SO(5)$ spinorial index.
A representation of the Gamma matrices is conveniently given by
\begin{equation}
\Gamma^{\mu}=\gamma^{\mu}\otimes 1, \quad
\Gamma^{5+i}={\tilde \gamma} \otimes \gamma^{i},
\end{equation}
where $\mu$ and $i$ are respectively  six and five 
dimensional vector indices, and ${\tilde \gamma}$ is the chirality matrix
in six dimensions:
\begin{equation}
{\tilde \gamma}= \gamma^{0}\dots\gamma^{5}.
\end{equation}
We shall also denote $\Gamma^{5+i}$ by $\Gamma^i$.
In this representation the  charge conjugaison matrix $C$ 
may be written as
\begin{equation}
C={\cal C}\otimes\Omega,
\end{equation}
where $\cal C$ is symmetric and verifies
\begin{equation}
{\cal C}^{-1}\gamma^{\mu}{\cal C}=-\gamma^{\mu\ T},
\end{equation}
whereas $\Omega$ is antisymmetric and verifies
\begin{equation}
\Omega^{-1}\gamma^i\Omega=\gamma^{i\ T}.
\end{equation}
Spinorial indices of $SO(5,1)$ and $SO(5)$ can be raised and
lowered with repctively $\cal C$ and $\Omega$.
The reduction of the algebra (\ref{onze}) to six dimensions leads to 
the (2,2) algebra
\begin{eqnarray}
\{Q^{+},Q^{+}\} & = & 2\Pi^{+}\gamma^{\mu}P_{\mu}\Pi^{+},\\
\{Q^{-},Q^{+}\} & = & 2\Pi^{-}\gamma^{i}c_{i}\Pi^{+},\\
\{Q^{-},Q^{-}\} & = & 2\Pi^{-}\gamma^{\mu}P_{\mu}\Pi^{-},
\end{eqnarray}
where the five-dimensional momentum appears as a central charge,
and $\Pi^{\pm}$ are the projectors on fermions of a given
six-dimensional chirality.
The  (2,2) algebra is invariant under the transformation
$Q^{-}\rightarrow -Q^{-}$, $c\rightarrow -c$. Modding out
with respect to this symmetry leads to the desired (2,0)
algebra. This is obtained by setting $Q^{-}=c^{i}=0$ in the above
formulae. As is evident from its construction this algebra
has a $Spin(5)=Sp(2)$ R-symmetry. The 11D Majorana fermion
becomes 
a $Spin(5)$ Majorana-Weyl six dimensional fermion :
\begin{equation}
\Psi=C\bar\Psi^{T},
\end{equation}
which reads in components
\begin{equation}
\Psi_{ a }=\Omega_{ a  b  }
{\cal C}\bar\Psi^{ b  \ T}.
\end{equation}
The antisymmetrised product of $n$ gamma
matrices is denoted  by
$\Gamma^{\hat\mu_1\dots\hat\mu_n}$. We have
\begin{equation}
(\Gamma^{\hat\mu_1\dots\hat\mu_n}C)^{T}=-(-1)^{n(n+1)/2}
\Gamma^{\hat\mu_1\dots\hat\mu_n}C,
\end{equation}
from which we get
\begin{eqnarray}
(\gamma^{\mu_1\dots\mu_{n}}{\cal C})^{T}
& = & 
(-1)^{n(n+1)/2}
\gamma^{\mu_1\dots\mu_{n}}{\cal C},\\
(\gamma^{i_1\dots i_{n}}\Omega)^{T}
& = & 
-(-1)^{n(n-1)/2}
\gamma^{i_1\dots i_{n}}\Omega.
\end{eqnarray}
A useful relation is the Fierz rearangement formula which
reads for four Weyl-Majorana fermions \cite{Fierz}
\begin{eqnarray}
&&\left( \bar\epsilon_1\Pi^{+}\epsilon_2 \right)
\left( \bar\epsilon_3\Pi^{+}\epsilon_4 \right)
=
\nonumber \\
&&
\hskip1cm
-\sum_{\substack{n_1=0,2\\n_2=0,1, 2}} 
{\frac{2}{c_{n_1}\tilde c_{n_2}}}
\left( \bar\epsilon_1
\Gamma^{\mu_1\dots\mu_{n_1}}
\Gamma^{i_1\dots i_{n_2}}\Pi^+\epsilon_4 \right)
\left( \bar\epsilon_3
\Gamma_{\mu_1\dots\mu_{n_1}}
\Gamma_{i_1\dots i_{n_2}}\Pi^+\epsilon_2 \right)
,\nonumber\\
&&
\left( \bar\epsilon_1 \Pi^{+}\epsilon_2 \right)
\left( \bar\epsilon_3\Pi^{-}\epsilon_4 \right)
 = 
\nonumber\\
&&
\hskip1cm
\sum_{n_2=0,1,2} 
{\frac{(-1)^{n_2}}{8\tilde c_{n_2}}}
\left(
-2 \left( \bar\epsilon_1\Gamma^{\mu_1}\Gamma^{i_1\dots i_{n_2}}
\Pi^- \epsilon_4 \right)
\left( \bar\epsilon_3\Gamma_{\mu_1}\Gamma_{i_1\dots i_{n_2}}
\Pi^+ \epsilon_2 \right) \right.
\nonumber\\
&&
\hskip1cm
+\left. \left( \bar\epsilon_1\Gamma^{\mu_1\mu_2\mu_3}
\Gamma^{i_1\dots i_{n_2}}\Pi^-
\epsilon_4 \right)
\left( \bar\epsilon_3\Gamma_{\mu_1\mu_2\mu_3}
\Gamma_{i_1\dots i_{n_2}}
\Pi^+\epsilon_2
\right)
\right)
\end{eqnarray}
with coefficients $c_{n}$ and $\tilde c_{n}$ given by
\begin{eqnarray}
c_{n} = 8\, (-1)^{n(n-1)/2} \ \mbox{ and } \
\tilde c_{n} =  4 \, (-1)^{n(n-1)/2}.
\end{eqnarray}
Using the Fierz rearangement  for commuting spinors $E$ and $F$, we get the 
useful
relations
\begin{equation}
(\bar E \Gamma_{\mu i}E)(\bar E\Gamma^\mu E)=0
,\label{re1}
\end{equation}
and
\begin{equation}
(\bar F\Gamma_{\mu\nu}E)(\bar E\Gamma^\nu E)
+(\bar F\Gamma^{i}E)(\bar E\Gamma_{\mu i} E)=0.\label{re2}
\end{equation}
\newpage


\end{document}